# Microscopic evidence of dominant excitonic instability in Ta$_2$NiSe$_5$

Seokjin Bae[1,2], Arjun Raghavan[1,2], Irena Feldman[3], Amit Kanigel[3], and Vidya Madhavan[1,2*]

[1]Department of Physics and Materials Research Laboratory, University of Illinois Urbana-Champaign, Urbana, Illinois 61801, USA
[2]The Grainger College of Engineering, University of Illinois Urbana-Champaign, Urbana, IL 61801, USA
[3]Department of Physics, Technion–Israel Institute of Technology, Haifa 32000, Israel

**An excitonic insulator (EI) is a charge-neutral bosonic condensate of spontaneously formed electron-hole pairs. Exotic quantum phenomena such as dissipationless charge neutral transport and huge potential for optoelectronic applications have led to an extensive search for intrinsic bulk materials that are EIs at ambient pressure without sophisticated device fabrication. The narrow gap semimetal Ta$_2$NiSe$_5$ has been proposed as a rare example of an intrinsic EI, but its ground state has remained controversial since the EI phase transition is accompanied by a structural distortion. Here, we use scanning tunneling microscopy and spectroscopy to present microscopic evidence that supports an excitonic origin of the insulating phase in Ta$_2$NiSe$_5$. First, we find that the insulating gap persists at structural domain boundaries where the bulk structural distortion is absent suggesting that the structural distortion is not the primary origin of insulating ground state. Second, the insulating gap is suppressed at localized charge puddles which indicates that charge correlations are important in producing the insulating gap. Finally, the decay length of the in-gap states at the puddles shows similar value to the estimated size of the exciton wavefunction from previous photoemission studies. Taken together, these findings suggest that Ta$_2$NiSe$_5$ is an EI and may be a versatile platform to study the physics of bosonic condensates in solid-state system at ambient conditions.**

*Correspondence to: vm1@illinois.edu

For a small gap semimetal or semiconductor where the carrier density is low, the Coulomb interactions between electrons and holes are poorly screened and spontaneously electron-hole bound pairs may form. When the binding energy exceeds the gap energy scale, the pairs condense into a macroscopic coherent state; dubbed as the excitonic insulator (EI)[1]. With close resemblance to both Bose-Einstein condensate and Bardeen-Cooper-Schrieffer type superconductor depending on the interaction strength[2–4], this bosonic condensate is expected to show exotic quantum phenomena such as dissipation-less transport[5,6] and superradiance[7]. Also, EIs have a demonstrated potential for variety of applications such as photodetectors[8], solar cells[9] and lasers[10].

Despite the interest in EIs, the identification of an intrinsic EI in a bulk material at ambient conditions has remained elusive. Systems where the presence of an EI state is unambiguously established are often fabricated devices[11–15,5,16–18] or in some cases, materials at high pressures[19,20]. The main difficulty in unambiguously identifying a material as an EI arises from the fact that the spectral gap in the insulating ground state can be caused not just by excitonic interactions, but also by other degrees of freedom that often coexist in a bulk system. For example, in one of the EI candidates 1$T$-TiSe$_2$[21,22], the presence of a charge density wave (CDW) possibly driven by band-type Jahn-Teller effect[23] hinders one from an unequivocal identification of an exciton-driven insulating ground state.

The narrow gap semimetal Ta$_2$NiSe$_5$ (TNSe) has been identified as a promising candidate bulk EI[24–29]. TNSe undergoes a semimetal to insulator transition at 326 K[30,31]. Angle-resolved photoemission (ARPES) studies report a low temperature spectral gap with a flat valence band top[32–36], one of the expected signatures of an excitonic insulator. While the material has no other competing electronic instabilities such as a CDW which can also open a spectral gap, the semimetal-to-insulator phase transition in this material is accompanied by an orthorhombic to monoclinic structural distortion[30,37,38] which may also account for the insulating ground state with a simple single-particle band gap opening.[34,39–43]. To distinguish between these two possibilities, one has to decouple the structural distortion from the excitonic interactions.

In this work, we employ scanning tunneling microscopy (STM) and spectroscopy (STS) to study single crystals of TNSe. We locate domain boundaries where the structural distortion is found to be absent and discover that the insulating state persists. This implies that the structural distortion is not the primary driver of the insulating ground state. In contrast, we find that local regions of excess charge (charge puddles) strongly suppress the insulating gap and that the decay length of the in-gap state at the charge puddles are in excellent agreement with the previously estimated size of the exciton wavefunction from ARPES[36]. These observations are consistent with a scenario where the locally enhanced carrier density at the charge puddles results in a suppression of the excitonic ground state over the size of the coherence length. Taken together, the STM data provide evidence of a dominant excitonic instability origin for the insulating ground state of TNSe, renewing the possibility of TNSe being a promising candidate for an intrinsic EI at ambient conditions.

The crystal structure of TNSe consists of quasi-1D chains of Ta, Ni, and Se atoms (Fig. 1a) along the a-axis. As shown in Fig. 1a, two Ta chains and one Ni chain are surrounded by five Se chains, forming one chain-unit. Adjacent chain-units along the c-axis, are inverted and have a height difference, leading to an alternating pattern of 'up Se chains' and 'down Se chains' (Fig. 1a). This plane of chain-units is stacked along the b-axis via van der Waals stacking. The sample cleaves along the a-c plane. The topography shown in figure 1 shows the Se chains with the 'up Se chains' being more prominent.

The high temperature phase (Fig. 1b), is orthorhombic and semimetallic[34,44]. Below the semimetal-insulator transition temperature of 326 K (Fig. 1c), the material undergoes a structural transition to a monoclinic phase[30,37,38] and concomitantly transitions into an insulator[34,35]. In the monoclinic phase, the two adjacent Se chains shear in the opposite directions (Fig. 1d). The Ta chains (which are not visible in the topography) follow suit albeit sheering in the opposite direction (Fig. 1c). At the same time, the electron-like band from the Ta chain (red band in Fig. 1b, c) and the hole-like band from the Ni chain (blue band) hybridize and open up a gap. Consistent with this scenario, tunneling spectra taken at 77 K show an insulating gap with a size of $\approx$ 160 meV (Fig. 1e), which is in agreement with the result from previous optical[25,31], and tunneling spectroscopies[45,46].

In the quest to answer whether this insulating transition is driven by the electron-hole interactions or the monoclinic shear distortion, the central bottleneck lies in the fact that macroscopically, the potential EI phase cannot be decoupled from the structural distortion. However microscopically we can find a detour around this problem. Since the structural distortion is a spontaneous broken symmetry, two types of domains with clockwise (CW; where the chain on the right moves down) and counter-clockwise (CCW; where the chain on the right moves up) distortions may exist in the sample. Indeed, by scanning different areas of the sample we find regions with both CW and CCW domains and a domain boundary between them (Fig. 2a). As seen from the enlarged topographic images in Fig. 2b and c, unlike the domains, the Se chains at the boundary show an absence of a clear shear distortion. To examine the relation between the structural distortion and insulating state, several STS linecuts along the chain direction (arrows in Fig. 2b, c) were obtained. As seen in Fig. 2d, e, both the domains and the boundary show a robust insulating gap whose size is independent of the amount and direction of the shear. To further verify this, STS linecuts across the chain direction (sky blue path in Fig. 2b, c) were obtained across the boundary and also show a consistent gap size (Fig. 2f, g). The observed robustness of the insulating state even in the absence (or marginal if any) of the structural distortion at the boundary implies that there may be another cause for the insulating ground state, i.e., an electron-hole interaction driven excitonic instability.

To obtain further insights into the potential EI phase in TNSe, we examine the effect of electron-hole interactions on the insulating state by studying areas with different localized electrostatic environments at local charge puddles. The increased carrier density at a charge puddle can provide enhanced screening of the Coulomb interaction between electron and holes, which could weaken the excitonic instability. Similar to magnetic impurities or vortices in a superconductor, studying the response of the system to charge puddles could provide important information on the EI order parameter.

Figure. 3a shows a large area topographic image showing different regions with excess charge. We focus here on charge puddles associated with subsurface defects which are ideal candidates to study the effects of a local electrostatic potential and the associated local charge on the pristine, defect-free surface layer. As we show next, type 1 and type 2 defects are precisely such defects occurring below the Se surface. We first consider the charge puddle is shown in Fig. 3b. The topography at a positive bias voltage (Fig. 3c) shows that the charge puddle is associated with an intact top surface chain indicating that it is indeed associated with a subsurface defect. Other examples of type 1 and type 2 defects are shown in Supplementary Fig. 1b. As a contrast, top-surface defects which are visibly associated with vacancies at different sites can be seen in Supplementary Fig 1c. The dI/dV spectra and maps on the subsurface defect (locations of the spectra and maps are displayed by stars and a yellow box in Fig. 3b) show semi metallic density of states localized to the defect, in contrast to the gapped insulating dI/dV spectra of its surrounding (Fig. 3d,e).

A trivial possibility for the enhanced density of states is that the subsurface impurity creates bound states within the gap. A more informative scenario is that the gap collapses due to the excess charge which can strongly screen the Coulomb interactions and locally collapse the EI phase. To distinguish between these scenarios, we study spatial decay lengths of the in-gap metallic states. For this purpose, two dI/dV linecuts are measured across the charge puddle (Fig. 4a). Figure 4b and c show the measured linecuts across (b) and along (c) the chain direction, showing metallic in-gap states peaking at the center of the puddle. Plots of the spatial profiles of the dI/dV along these linecuts at an energy inside the gap (-130 mV) show that both profiles fit well to the Gaussian (Fig. 4d, e) with 1/e decay diameter $2\xi_c$ = 0.49 nm (across the chain, c-axis) and $2\xi_a$ = 1.57 nm (along the chain, a-axis).

These quantitative observations suggest the scenario depicted in Fig. 4f. The extra height at the puddle implies an extra carrier density which can suppress the excitonic interaction via enhanced screening. Then, analogous to a superconductor with local magnetic impurities, one can expect the excitonic order $\psi$ at the puddle to recover within a distance determined by the size of the electron-hole pair wavefunction $2\xi$. Indeed, the estimated decay diameters ($2\xi_a$, $2\xi_c$ = 1.57 nm, 0.49 nm) are in very good agreement with the estimated size of the exciton wavefunction from a previous photoemission study

($2\xi_a$, $2\xi_c$ = 1.5 nm, 0.6 nm)[36]. Also, the across-the-chain decay radius $\xi_c$ = 0.245 nm is similar to the distance between Ta and Ni chains (0.22 nm) where each low energy electron and hole resides[34,45,47]. This coincidence would be natural if the suppression of insulating state is mediated by the breaking of electron-hole pair between Ta and Ni chains as illustrated in Fig. 4g. and Ref.[27,29,36,47]. The quantitative comparison of the decay lengths of the semi metallic state to the previously known system parameters supports the scenario that excitonic instability is the main origin of the insulating ground state. This quantitative analysis is reproducible as shown in (Table. 1).

In summary, using STM imaging and spectroscopy, we have been able to decouple the structural transition from the electronic ground state which has been a long-standing hurdle in the debate on whether $Ta_2NiSe_5$ is an excitonic insulator. Our results show that the insulating ground state of $Ta_2NiSe_5$ is robust even when the structural shear distortion is significantly reduced; that the insulating state can be suppressed by enhanced local carrier density; and that the suppression diameter is related to the size of the exciton wavefunction. Our findings reveal evidence for the marginal role of the structural distortion and the dominant role of the excitonic instability on the insulating phase of $Ta_2NiSe_5$.

## Methods

Single crystals of $Ta_2NiSe_5$ family were grown by chemical vapor transport method[27]. Elemental powders of tantalum, nickel, and selenium were mixed with stoichiometric ratio and then sealed in an evacuated quartz ampule with a small amount of iodine as the transport agent. The mixture was placed in the hot end of the ampule (~950∘C) under a temperature gradient of about 10∘C/cm. After about a week mm-sized needle-like single crystals were found at the cold end of the ampule. The crystals are shiny and cleave easily. We used x-ray diffraction (XRD) and electron dispersive X-ray spectroscopy to verify the exact composition of the crystals and their uniformity. For the STM measurements, samples were cleaved at room temperature (295 K) in an ultra-high vacuum. After cleaving, samples are directly transferred to the STM. All STM measurements were performed at 77 K with annealed tungsten tips. STS spectra were collected standard lock-in technique at a frequency of 941 Hz.

**Acknowledgements**

The STM work is supported by Gordon and Betty Moore Foundation's EPiQS initiative through grant number GBMF9465 and Q-NEXT, supported by the U.S. Department of Energy, Office of Science, National Quantum Information Science Research Centers. The authors are immensely grateful to Pavel Volkov for helpful discussions.

**Author Contributions:**

S. B., and V. M. conceived the project. S. B. and A. R. conducted STM measurement. I.F. and A. K. prepared the samples used in this study. S. B. and V. M. performed data analysis and wrote the manuscript with input from all the authors.

**Competing interests:** The authors declare that they have no competing interests.

**Data Availability:** Data for the main figures will be available at the Illinois Databank.

**Figure 1**

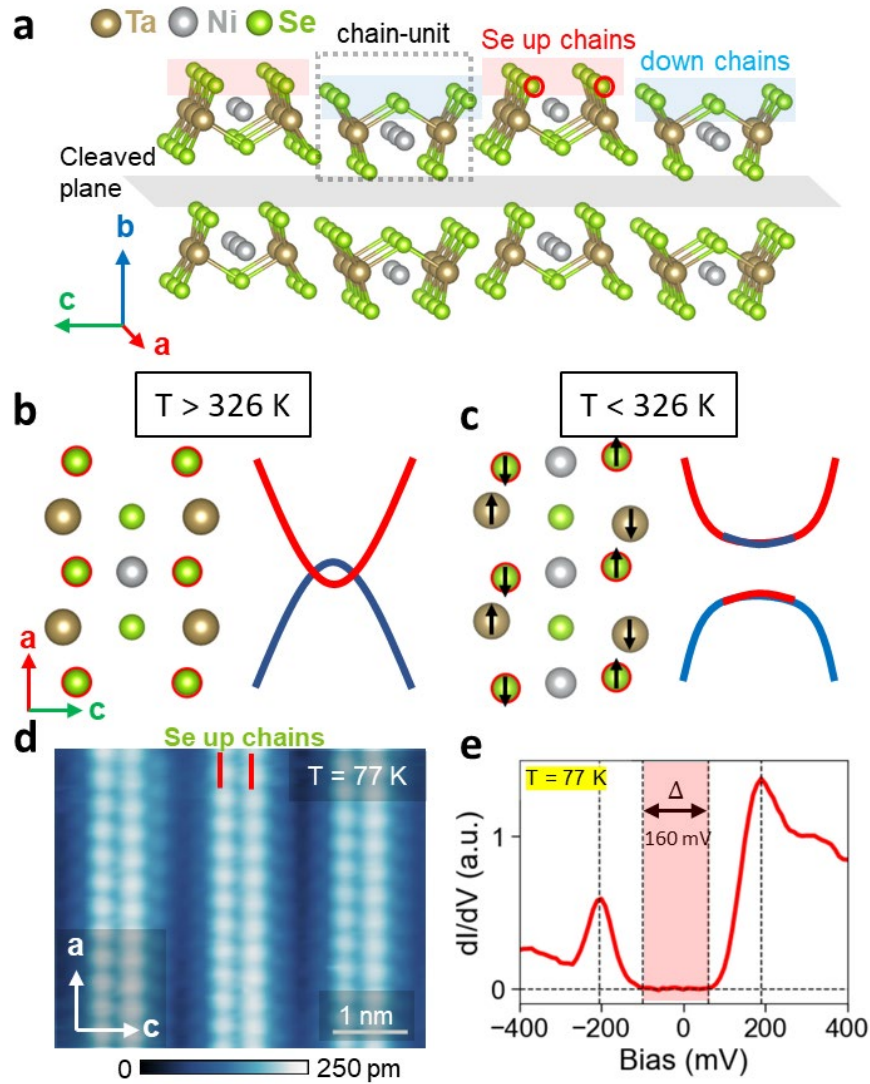

**Fig. 1: Crystal structure, phase transition, topography, and tunneling spectroscopy of Ta$_2$NiSe$_5$.**

**a**, Crystal structure and crystal axis of Ta$_2$NiSe$_5$. Cleaved plane is normal to the *b*-axis. The grey dotted box denotes chain-unit. Ta$_2$NiSe$_5$ forms a chain-like structure along *a*-axis with the height modulation between up-chains and down-chains. Two Se-atom chains (circled in red) are prominent in the topographic image as shown in (d). **b, c,** top view of the crystal structure and the schematic low energy band diagram above (b) and below (c) the phase transition temperature $T_s \approx 326$ K. The arrows in (c) depict the direction of displacements (exaggerated for clear display) in atomic positions under transition. **d**, An STM topographic image below $T_s$ ($V_{bias}$ = - 400 mV, $I_{set}$ = 50 pA), showing a shear distortion between two Se atom up-chains. **e**, A representative dI/dV spectrum which shows an insulating gap of $\approx 160$ mV.

**Figure 2**

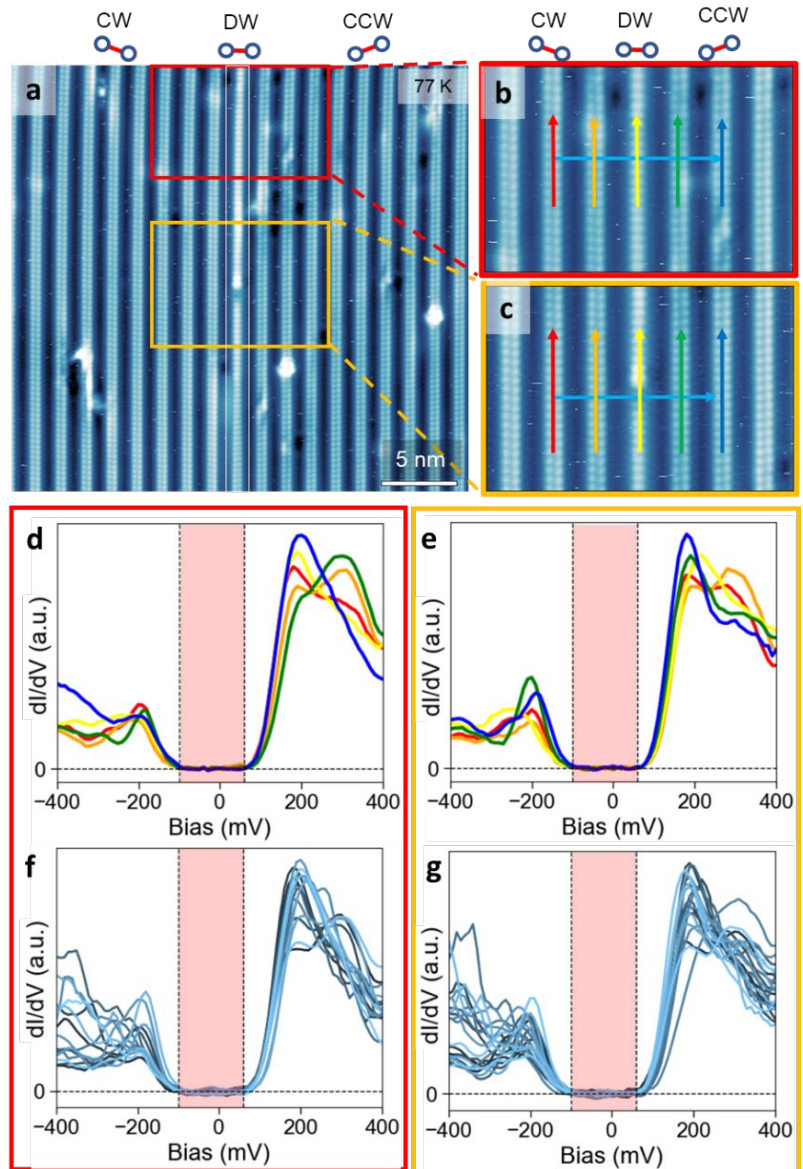

**Fig. 2: Robust insulating state at the domain boundary where the structural distortion is significantly reduced.**
**a**, A topographic image which shows a domain boundary (DW: white rectangle) between regions with opposite directions of shear distortion ($V_{bias}$ = - 450 mV, $I_{set}$ = 30 pA). The region on the left shows clockwise (CW) shear between two Se up-chains. The region on the right shows counter-clockwise (CCW) shear. The domain boundary shows marginal (if any) shear. **b, c**, A close-up of the area in the red (b) and orange (c) box in (a). Averaged dI/dV spectra along the five vertical arrows in (b, c) are shown in **d, e**. The color of the spectra matches with the color of each arrow. **f, g**, dI/dV linecut along the horizontal arrow (skyblue) in (b, c). All the spectra along and across the shear distortion domain boundary shows a robust insulating gap denoted as the red box in (d-g), implying the structural phase transition is not a dominant origin of the insulating ground state.

**Figure 3**

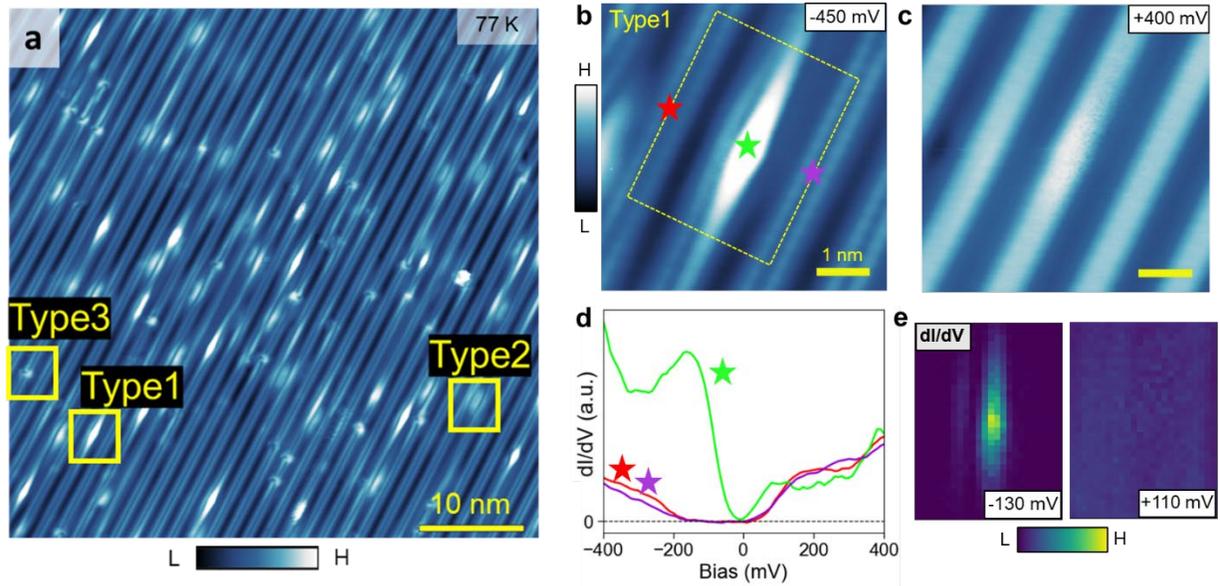

**Fig. 3: Suppression of the insulating state due to local electrostatic environment.**
**a**, a 50 x 50 nm² topographic image shows various types of local electrostatic environment ($V_{bias}$ = - 450 mV, $I_{set}$ = 30 pA). **b**, A topographic image of the type1 at negative bias ($V_{bias}$ = - 450 mV, $I_{set}$ = 30 pA) shows a local electrostatic environment with a charge puddle on a Se up-chain. Stars in each color represent positions where dI/dV spectra shown in (d) were taken. The yellow dashed box is the 3 x 4 nm² area where dI/dV maps in (e) are taken. **c**, A topographic image at positive bias ($V_{bias}$ = 400 mV, $I_{set}$ = 30 pA) shows intact lattice at the top surface, implying the puddle feature shown in (b) is not from a top-surface defects such as adatom, substitution, or vacancy. **d**, The lime (red, violet) star represents a position on (away from) the local electrostatic environment. The localized charge puddle without top surface defects shows semimetallic spectra, implying a suppression of the insulating state. **e**, dI/dV maps around the charge puddle (yellow dashed box in (b)) at energies within the insulation gap (left, -130 mV) and outside the gap (right, +110 mV).

Figure 4

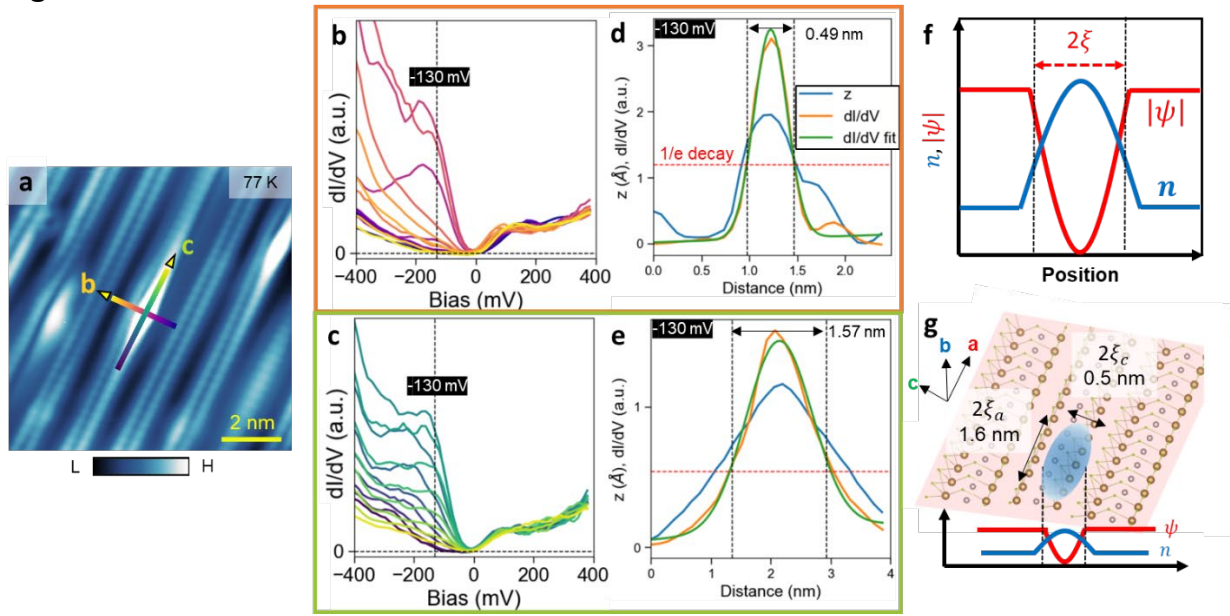

**Fig. 4: Decay length of the semimetallic states at a charge puddle provides an estimate for the size of exciton wavefunction.**
**a**, A 10 x 10 nm² topographic image shows a localized charge puddle (white ellipse, same spot as Fig. 3b) ($V_{bias}$ = - 450 mV, $I_{set}$ = 30 pA). The lines with plasma and viridis colormap denote the path for dI/dV linecut through the charge puddle across the chain (b) and along the chain (c) directions. **b**, **c**, The dI/dV linecuts along the paths shown in (a). **d**, **e**, The topographic height (blue), dI/dV at - 130 mV (orange) which represents states within the insulation gap, and Gaussian fit (green) to the dI/dV along the linecuts in (b), (c). The fit to dI/dV shows 1/e decay length of 0.49 nm across the chain (c-axis) and 1.57 nm along the chain (a-axis) direction. **f**, A schematic relation between local carrier density and excitonic order parameter. On a position where local carrier density ($n$) is high, the enhanced screening of Coulomb interaction between electrons and holes can introduce suppression of the excitonic (insulating) order ($\psi$), resulting in semimetallic state at the puddle. $\xi$ is the radius of the exciton wavefunction. **g**, Illustration of the bubble where the excitonic order is suppressed (blue) by a charge puddle, overlayed on the lattice. The size of the bubble represents the size of the exciton wavefunction estimated by the decay length of the in-gap semimetallic state in (d) and (e).

Table 1

|  |  | $2\xi_a$ (nm) | $2\xi_c$ (nm) |
|---|---|---|---|
| up-chain puddle | Type1 spot1 | 1.57 | 0.49 |
| up-chain puddle | Type1 spot2 | 1.63 | 0.76 |
| down-chain puddle | Type2 spot1 | 1.98 | 0.58 |
| down-chain puddle | Type2 spot2 | 1.83 | 0.59 |
|  | Estimated wavefunction size (ARPES)[36] | 1.5 | 0.6 |

**Table. 1: Decay diameter of the in-gap metallic states along the a-axis and c-axis.**

**Supplementary Figure 1**

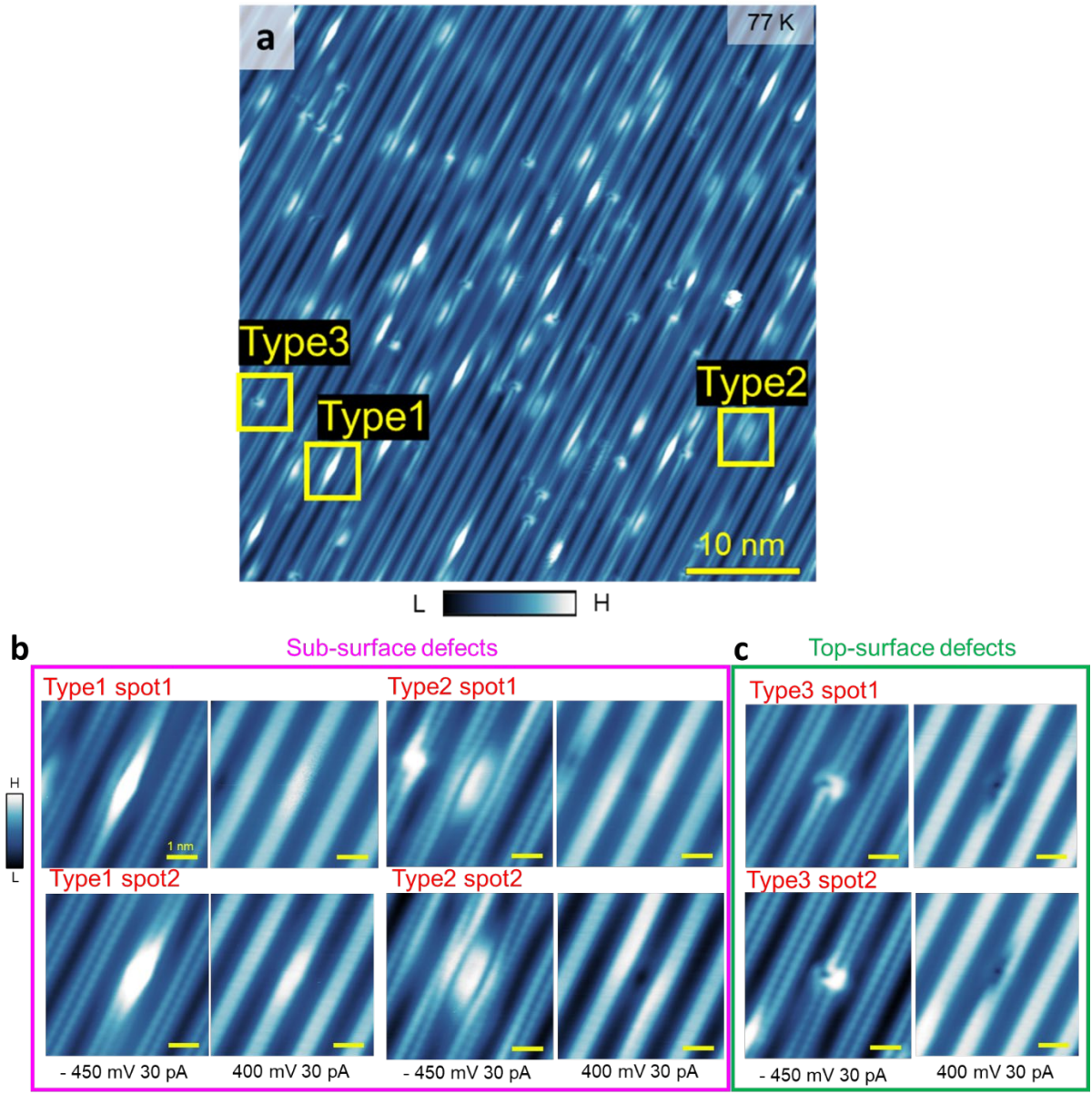

**Supplementary Fig. 1: Types of local electrostatic environments.**
**a**, a 50 x 50 nm$^2$ topographic image shows various types of local electrostatic environment ($V_{bias}$ = - 450 mV, $I_{set}$ = 30 pA) which is also shown in Fig. 3a of the main text. **b,c**, Close-up topographic images of each type of local electrostatic environments at positive and negative biases. **b**, Type1 and 2 (magenta box): a charge puddle on a Se up-chain and a down-chain due to a sub-surface defect. **c**, Type3 (green box): a Se vacancy on the top-surface.

## Supplementary Figure 2

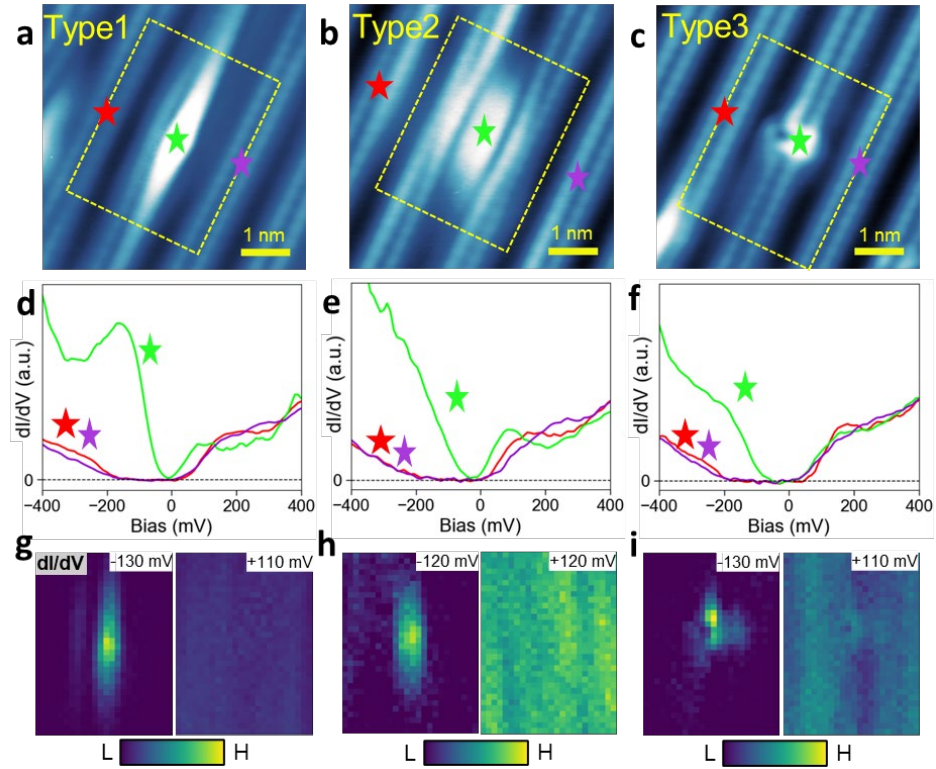

**Supplementary Fig. 2: Effect of each type of local electrostatic environments on the insulating ground state.**
**a-c**, Close-up topographic images of each type of local electrostatic environments ($V_{bias}$ = − 450 mV, $I_{set}$ = 30 pA). Stars in each color represent positions where dI/dV spectra shown in (d-f) were taken. The yellow dashed boxes are the 3 x 4 nm$^2$ area where dI/dV maps in (g-i) are taken. The lime (red, violet) star represents a position on (away from) the local electrostatic environments. **d,e,** Localized charge puddles without surface vacancies (type 1, 2) shows semimetallic spectra, implying a suppression of insulating state. **f**, A top-surface vacancy (type 3) shows a spectrum with smaller but still existing insulating gap. **g-i**, dI/dV maps around each type1-3 spot (yellow dashed boxes in **a-c**) at energies within the insulation gap (left) and outside the gap (right).